\documentclass[11pt,twoside]{article}

\usepackage{asp2006}
\usepackage{epsf}
\usepackage{graphicx}    

\begin{document}
\title{Invisible Major Mergers: Why the Definition of a Galaxy Merger Ratio Matters.}
\author{Kyle R. Stewart}
\affil{Center for Cosmology, Department of Physics and Astronomy, The University of California at Irvine, Irvine, CA, 92697, USA}

\begin{abstract} {The mapping between dark matter halo mass, galaxy stellar mass,
and galaxy cold gas mass is not a simple linear relation,
but is influenced by a wide array of galaxy formation processes.
We implement observationally-normalized relations
between dark matter halo mass, stellar mass, and cold gas mass to
explore these mappings, with specific emphasis on the correlation 
between different definitions of a major galaxy merger. 
We always define a major merger by a
mass ratio $m/M>0.3$, but allow the masses used to \emph{compute} this ratio
to be defined in one of three ways:
dark matter halo masses, galaxy stellar masses, or galaxy baryonic masses (stars and cold gas).
We find that the merger ratio assigned to any particular merger event depends
strongly on which of these masses is used, with the mapping between different
mass ratio definitions showing strong evolution with halo mass and redshift.
For example, major dark matter mergers ($>0.3$) in small galaxies
($M_{\rm DM}<10^{11}M_{\odot}$) typically correspond to very minor \emph{stellar} mergers
($<1/20$).  These mergers contain significant dark matter mass, and should cause 
noticable morphological disruption to the primary galaxy, even though there is no 
observable bright companion.  In massive galaxies, there is an opposite effect,
with bright companion galaxies corresponding to only minor dark matter mergers.
We emphasize that great care must be taken when comparing mergers based on 
different mass ratio definitions.
}

\end{abstract}

\section{Introduction}
\label{Introduction}
In the cold dark matter (CDM) model of structure formation, galaxy mergers
are believed to play an important role in galaxy evolution.
Typically, these mergers are divided into two categories.
``Minor'' mergers (with mass ratios $<1/3$)
are often thought to trigger moderate bursts of increased star formation and/or
morphological disturbances, as well as contributing to the deposition of diffuse light components
of galaxies.  ``Major'' mergers (with mass ratios $>1/3$)
are likely to influence stronger morphological disturbances responsible for the transformation
from disk-dominated to bulge-dominated morphologies, in addition to triggering stronger starburst
and AGN activity.

Despite this commonly adopted distinction
between major and minor mergers at merger mass ratios of $\sim1/3$, there is still ambiguity in 
what mass is used to \emph{define} this ratio.
Theoretical investigations of dark matter halo
merger rates typically define merger ratios in terms of dark matter halo masses,
the most theoretically robust prediction from cosmological $N$-body simulations
\citep[e.g.][and references within]{Stewart08}.
But because estimates of dark matter halo masses are difficult to obtain observationally,
it is also common to define merger ratios by comparing the stellar masses or the total
baryonic masses of galaxies.

In attempting to compare theoretically derived merger statistics 
(in terms of dark matter mass ratios) to
observational investigations of galaxy mergers
(in terms of stellar or galaxy mass ratios), it is important to understand
the mapping between these definitions.  Galaxy merger rates, for example,
are quite sensitive to the merger mass ratio being considered \citep{Stewart08b}.
In order to explore the fundamental differences between
major mergers as defined by dark matter halo, stellar, and galaxy merger ratios,
we adopt a semi-empirical methodology to estimate the stellar and cold gas content of
dark matter halos as a function of halo mass and redshift.  
We give a very brief overview of this method before presenting our 
findings, but we refer reader to \cite{Stewart09b}
for a more in-depth discussion of this method.

\section{Assigning Baryons and Defining Masses}
In order to assign stars to our halos, we assume
a monotonic relationship between halo mass and stellar mass.
Using this technique, provided we know $n_g(>M_{\rm star})$ (the cumulative number
density of galaxies with stellar mass more massive than $M_{\rm star}$)
we may determine the associated dark matter halo population
by finding the halo mass above which the number density of halos
(including subhalos) matches that of the galaxy population, $n_h(>M_{\rm DM}) = n_g(>M_{\rm star})$.
Specifically, we adopt the relation found by
\citealp{ConroyWechsler08} (interpolated from the data in their Figure 2).
Of course, a simple relation of this kind
cannot be correct in detail, but in an average sense, it provides a good characterization
of the relationship between halo mass and galaxy stellar mass that must hold in order for
LCDM to reproduce the observed universe.

In order to assign gas to the central galaxies within our halos, we
quantify observationally-inferred relations between gas fraction and stellar mass.
Specifically, we characterize the data from
\citealp{McGaugh05} (disk-dominated galaxies at $z=0$)
and \citealp{Erb06} (UV-selected galaxies at $z\sim2$) with a
relatively simple function of stellar mass and redshift \citep[see][]{Stewart09b},
and find that this adopted characterization is also consistent
with a number of other observationally motivated
works \citep[e.g.,][]{Kannappan04, Baldry08}.

Having estimated the stellar and cold gas content of
dark matter halos as a function of halo mass and redshift, we define three
different means of identifying the mass of a galaxy (and thus, define merger mass ratios):
\begin{enumerate}
\vspace{-1ex}
\item The mass (or mass ratio) of each dark matter halo, $(m/M)_{\rm DM}$.
We will refer to these as the \emph{DM mass (ratio)} of a galaxy (merger).
\vspace{-1ex}
\item The mass (mass ratio) of the stellar mass of each dark matter halo's central
galaxy, $(m/M)_{\rm star}$.  We refer to this definition as the \emph{stellar mass (ratio)}.
\vspace{-1ex}
\item The mass (mass ratio) of the total baryonic mass of each dark matter halo's central
galaxy, $(m/M)_{\rm gal}$.  In this case, we define a galaxy's baryonic mass
as a combination of its stellar mass and cold gas mass
($M_{\rm gal} \equiv M_{\rm star} + M_{\rm gas}$).  We refer to this definition
as the \emph{galaxy mass (ratio)}.
\end{enumerate}

\begin{figure*}[t!]
  \plotone{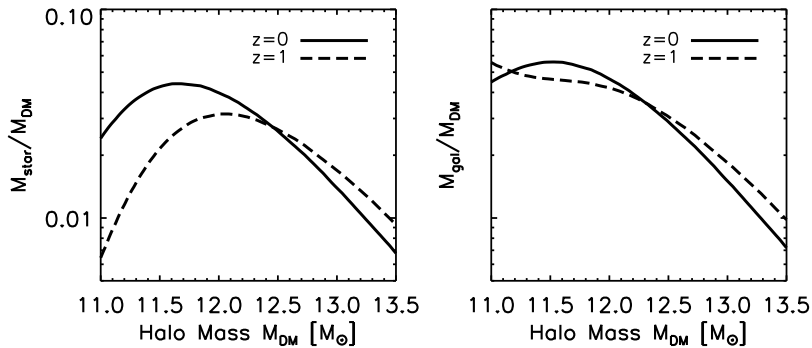}
    \caption{A comparison of the baryonic properties of central galaxies to their dark matter halo masses.
\emph{Left:} Ratio of stellar mass to dark matter halo mass, as a function of
halo mass.
\emph{Right:} Ratio of total baryonic mass (stars and cold gas) to halo mass, as a function of halo mass.
In both panels, the solid and dashed lines correspond to $z=0$ and $z=1$, respectively.  
Note that these relations vary significantly with halo mass and redshift.}
\label{MMdm}
\end{figure*}

Using these mass definitions, we show the
stellar and galaxy mass of a dark matter halo's central galaxy as a function of
halo mass (and normalized by halo mass) in Figure \ref{MMdm}, where solid and dashed lines represent
galaxies at $z=0$ and $z=1$, respectively.  
We emphasize that these mass fractions
are a strong function of halo mass, and evolve with redshift.
It is clear from this figure that a single merger event between  
galaxies may have a drastically different mass ratio
in dark mater compared to its mass ratio in stars (or baryons).     
This could have important implications for observational efforts to  
measure the merger rate: morphological disturbances will be affected  
by high total mass ratio events, while pair count estimates will be  
more sensitive to the mass ratio in visible light.

\begin{figure*}[t!]
  \plotone{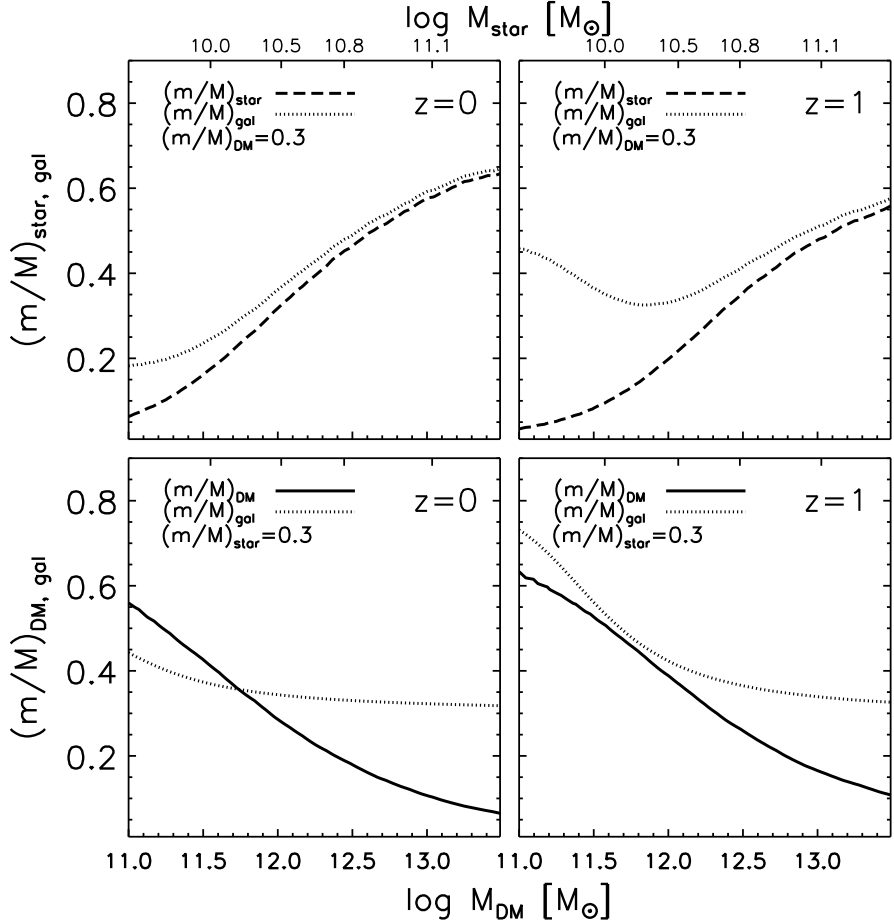}
  \caption{Conversion between major merger ratios defined by dark matter halo mass ($M_{\rm DM}$),
stellar mass ($M_{\rm star}$), or galaxy (baryonic) mass
($M_{\rm gal}$). In the top
panels, the dashed and dotted lines show the stellar and galaxy
mass ratios of major DM mergers with $(m/M)_{\rm DM}=0.3$, as a function of
halo mass (lower axis) and stellar mass (top axis).
Similarly, the solid and dotted lines in the bottom panels show the DM and galaxy
mass ratios corresponding to major stellar mergers with $(m/M)_{\rm star}=0.3$.  
The left and right panels give these relations at $z=0$ and $z=1$, respectively.}
\label{mMdef}
\end{figure*}

\section{Mapping Between Mass Ratios}
We explore the various mappings between major mergers using different mass definitions 
in Figure \ref{mMdef}.  In the top panels we focus on mergers with
$(m/M)_{\rm DM}=0.3$ (henceforth \emph{major DM mergers}),
with the dashed and dotted lines showing the corresponding stellar
and galaxy mass ratios of these mergers.  In the bottom panels, we instead focus on
\emph{major stellar mergers}, defined by $(m/M)_{\rm star}=0.3$.  In these panels, the solid
and dotted lines correspond to the DM and galaxy mass ratios of these mergers.  The
left and right panels show relations at $z=0$ and $z=1$, respectively.

We emphasize that, in general, the DM mass ratio
between two galaxies is \emph{not} the same as the stellar (or galaxy) mass ratio.
Specifically, major DM mergers (top panels) correspond to stellar mass ratios ranging from $\sim5-60\%$ $(5-50\%)$
and galaxy mass ratios of $\sim20-60\%$ $(35-50\%)$ for
$10^{11-13} M_{\odot}$ halos at $z=0$ ($1$).  Similarly, major stellar mergers (bottom panels)
correspond to DM mass ratios from $\sim10-55\%$ $(15-65\%)$ and galaxy mass ratios
of $\sim30-45\%$ $(35-75\%)$ for $10^{11-13} M_{\odot}$ halos
at $z=0$ $(1)$.  Indeed, the \emph{only} broad regime where different mass definitions
result in similar merger ratios is for stellar and galaxy merger ratios of
massive galaxies, where galaxy gas fractions are typically low enough that
$M_{\rm star}\sim M_{\rm gal}$.

Note that in $M_{\rm DM}<10^{11}M_{\odot}$ halos,
major DM mergers should contain stellar mass ratios $<1/20$.  The smaller galaxy 
in these mergers contain significant 
mass in dark matter, and should be capable of triggering severe morphological disruption
in the primary galaxy, but they are observationally ``invisible,'' 
with negligible luminous content with respect to the primary.  
The existence of these ``invisible''
major mergers is a robust, testable prediction of LCDM.

\section{Example Consequence: Measuring the Merger Rate}
While theoretical investigations into dark matter halo merger rates define mergers by DM mass ratios,
observed merger rates (specifically, those based on close-pair counts of galaxies) typically
select pairs based on luminosity (stellar mass), and should thus constitute major \emph{stellar} mergers.
The mapping between stellar and DM mass ratios has two important qualitative effects in this case.
First, for smaller halos, major DM mergers should 
correspond to substantially smaller stellar mass ratios, 
and may not be distinguishable as a luminous close-pair when observed (ie.~faint/invisible major mergers).
Second, for massive halos, some observed close-pairs with comparable luminosities
(major stellar mergers) may correspond to substantially smaller \emph{DM} mass ratios, and would be 
counted as \emph{minor} (not major) mergers in theoretical predictions from $N$-body simulations (ie.~bright minor mergers).

For a more quantitative analysis, 
we adopt the fitting function for $dN/dt$ from
\citealp{Stewart08b} (Table 1, infall, ``simple fit''),
which provides the rate of mergers more massive than $(m/M)_{\rm DM}$ into dark matter
halos of mass $M_{\rm DM}$ (per halo, per Gyr) as a function of redshift, mass, and mass ratio.  
For $10^{11}M_{\odot}$ dark matter halos and DM merger ratios $>30\%$ at $z=0-1$, 
the merger rate increases from $\sim 0.015 - 0.075$.
Now consider an identical halo mass, but for mergers selected on
\emph{stellar} mass ratios $>30\%$, corresponding to a DM mass ratio of 
$\sim7\%$ ($4\%$) at $z=0$ ($1$).  Because of the minor DM mergers being 
considered, this selection would result in an artificial increase 
of the observed merger rate by a factor of $3-4$ compared to DM merger rates 
($\sim0.05 - 0.30$ from $z=0-1$), with a redshift evolution that is too steep
\citep[and does not appear to fit well 
to $dN/dt \propto (1+z)^{\alpha}$; see][]{Stewart08b}.
Thus, using different mass ratios to define mergers has a substantial effect on 
predictions of galaxy merger rates, and 
great care must be taken when comparing studies of galaxy or halo mergers, if the
merger mass ratios have been defined by different criterion.

\acknowledgements KRS thanks Shardha Jogee and the conference organizers
for an excellent conference with a thought provoking program.

\end{document}